
\magnification=1200
\baselineskip 18pt
\parskip 3pt
\centerline{THE OPTICAL GRAVITATIONAL LENSING EXPERIMENT.}
\centerline{THE EARLY WARNING SYSTEM.}
\vskip 1cm
\centerline{A. Udalski\footnote{$^*$}{Warsaw University Observatory, Al.
Ujazdowskie 4, 00--478 Warszawa, Poland}$^,$\footnote{$^{\dag}$}{Princeton
University Observatory, Princeton, NJ 08544}, M. Szyma\'nski$^{*,\dag}$, J.
Ka\l u\.zny$^{*,\dag}$, M. Kubiak$^{*,\dag}$,}
\centerline{M. Mateo\footnote{$^{\ddag}$}{Department
of Astronomy, University of Michigan, 821 Dennison  Bldg., Ann Arbor,
MI 48109--1090}, W. Krzemi\'nski\footnote{$^{\|}$}{Carnegie Observatories, Las
Campanas Observatory, Casilla 601, La Serena, Chile}, B. Paczy\'nski$^{\dag}$}
\vskip 1cm

{\bf

The discoveries of 17 microlensing event candidates have been reported
over the last year by three teams conducting unprecedented mass
photometric searches in the direction of the Galactic bulge and the
Magellanic  Clouds.  These include 10 events found by the OGLE
collaboration$^{1,2,3,4}$, 5 by the MACHO team$^{5,6}$ and 2 by  the
EROS team$^{7}$. All searches have the main goal to detect dark matter
in our Galaxy.  The detection of 17 event candidates proves that the
microlensing is a powerful tool in the search for dark matter$^{8,9}$,
and it may be used for reliable mass determination when the geometry of
the event is known. Here we present the first microlensing event,
OGLE~\#11,  discovered in real time, using the newly implemented "Early
Warning System". We describe our system which makes it possible to
monitor and study in great details any very rare phenomena, not only
lensing events, with a broad array of instruments almost immediately
after they have changed their brightness.

}

The probability of occurrence of gravitational  microlensing is very
small even in the dense stellar fields of LMC or the Galactic bulge. The
first estimate of the optical depth to gravitational microlensing
towards the Galactic bulge is $ \sim ( 3.3 \pm 1.2 ) \times 10^{-6}$ $
^3$. Thus observations of millions  stars conducted over a period of
years are necessary to collect a significant sample of events.  The
OGLE  project has conducted its search in the direction of the Galactic
bulge since 1992  using the  Swope 1-m telescope at Las  Campanas
Observatory, Chile,  which is operated by Carnegie  Institution of
Washington.  A  $2048\times2048$ Loral  CCD chip is used as the
detector.  Currently, about 6 millions photometric  measurements of 4
millions bulge objects are collected  during each clear night.

The characteristic feature of such a massive survey is the  tremendous
data rate, and the greatest difficulty is the need to process the data
quickly.  Typically, the  reductions are usually significantly delayed.
Up to now, all microlensing events were announced well after the events
had occurred, with delays  ranging from  3 months (OGLE~\#1) to almost
3~years (EROS~\#2).  All the lensed stars were back at their normal
brightness when  the actual discovery (as opposed to the observations)
was made.  This delay also introduces an  important practical problem:
the light curves of most previous microlensing events have non-uniform
or sparse temporal coverage.  This is partly due to  uncontrollable
factors such as poor weather or  telescope scheduling (the latter is a
particularly major limitation for OGLE which does not have a dedicated
telescope).  However, in many cases better data could have been obtained
if the observers had known that some particular fields contained
promising candidate events.  Thus it became evident that we have
to develop a system  which would allow the  detection of a microlensing
event soon after it starts. Such an event could then be widely observed
in all possible wavelength ranges photometrically as well as
spectroscopically from many observatories around the world. Crucial
parts of the light  curve near maximum light could be covered with a
very good time  resolution.  These parts of the microlensing light curve
 provide potential information about binarity of lensing star or  even
about possible planets around it$^{10}$.   Also, good coverage of the
caustic crossings in double microlens events may  provide important data
about the structure of the lensed objects, and the determination of
parameters of the binary lensing system$^4$.

Beginning with the 1994 observing season (March-September) the OGLE
project has implemented an "Early Warning System" (EWS) developed at
Warsaw University Observatory$^{11}$.  This software system is designed
to detect on-going microlensing events and it opens a new era in massive
photometric surveys by allowing early detection of microlensing
phenomena and  follow-up observations from other observatories.  The
system was designed to  detect any kind of transient and smooth stellar
variability, not just microlensing.

The microlensing detection technique used by OGLE was  described in
details in a series of papers$^{1,2,3}$.  To  summarize, a database of
about 2 million non-variable stars in 1992-93 seasons was constructed
from the original database of all measured objects.  All photometry was
performed using a modified version of the DoPhot photometry
program$^{12}$. On average, about 30 CCD frames of the Galactic bulge
are collected during a night. The computer system consisting of
multiprocessor Sparc 10 workstation is capable  of reducing the data
from a typical night within 24 hours.  The new measurements  of each
object  from the non-variable stars database is compared with its
database mean  magnitude. If the brightness differs more than a certain
threshold  appropriate to the  mean magnitude of the star, the star is
flagged by the EWS as  suspected.  Next, the EWS checks whether the
brightness change is continuous. If the  number of consecutive
measurements giving the brightness above the threshold exceeds some
fixed number then an alarm is issued.   The threshold is set at three
times the maximum allowed sigma of brightness for a non-variable star of
given magnitude, and the number of consecutive observations above the
threshold is set to five -- a compromise between number of false  alarms
and detection delay.  Both parameters can be  tuned as needed depending
on the type of variables one is searching for. Alarms are limited to
objects which increase their brightness (like microlenses) but can
easily  be extended to include the objects whose magnitude steadily
drops.

The light curves of the objects signalled by the EWS, are automatically
e-mailed to the OGLE headquarter at the Warsaw University Observatory.
Objects are  further analyzed there to  exclude  obvious false alarms
due to bad photometry, CCD detector  blemishes etc.  A variety of
filters which reject random variability can be applied at that stage.
For example, the errors of  observation which are  sensitive indicators
of the quality of photometry,  are analyzed. If the  suspected object
passes this step then visual inspection of the CCD frame is performed to
check whether the photometry is not affected by CCD defects.  Next, the
light curve is analyzed for the microlensing type of light variability.
If all tests are positively passed and brightening of the object is
confirmed by two additional observations, the suspected  object is named
a "prime microlens candidate" and a world-wide alarm  is distributed to
the astronomical community.

The EWS system had been intensively tested on real data from  the
previous seasons prior to its implementation. All earlier OGLE
microlensing events were detected at the early stages of their
developement with no false alarms.  The EWS system has been routinely
used in the 1994 season.   Soon  after its implementation, a new
microlensing candidate has been discovered. The object: BW6 Baade's
Window field star$^{13}$ I 167045, was detected by the EWS in July 1994.
The star passed all additional criteria and  a world wide alert was
issued.  The J2000.0 coordinates of the star are
$\alpha_{2000}=18^h03^m45^s.1$, $\delta_{2000}=-30^{\circ}18'16''.4$ and
$66''\times 66''$ region  of the CCD frame around OGLE~\#11  is
shown in Fig.\ ~1.

As further data were obtained, we confirmed that the star  is indeed a
good microlensing candidate.  Its  light curve  (complete through date)
and the best  fit to the theoretical  microlensing curve is shown in
Fig.\ 2.  The solid horizontal line  shows the mean 1992-93 magnitude,
and the dashed line  shows the threshold level.    The vertical arrow
points to the time of discovery.  The event has been designated as
OGLE~\#11.   This event, OGLE~\#11,  is the first microlensing event
detected while it was still unfolding.

At the moment of discovery it was not clear whether the  event is still
on its rising part of the light curve. It  turned out, however, that
OGLE~\#11 is a low amplitude event  and it reached the EWS threshold,
and was discovered near the maximum brightness.   Nevertheless the event
could be followed during the descending branch of the light curve. The
best-fit parameters of the OGLE~\#11 light curve are:  time of maximum
brightness, $T_{max}\hbox{(JD hel.)}=2449537.3 \pm 0.7$; time scale (the
Einstein radius / transverse velocity), $t_0=12.6\pm 1.3$~days; magnification,
$A=1.32\pm 0.02$;  normal $I$ magnitude,~ $I_0=18.22\pm 0.01$; ~ $\chi^2 ~
\hbox{(dof)}=0.72$.

The normal colors of OGLE~\#11 are $V=19.7$, $V-I=1.6$  indicating a
normal Galactic bulge turn-off point main sequence  star.  OGLE~\#11 is
one of the smallest amplitude events ever  detected; its discovery
dramatically proves the efficiency of the EWS system.

The EWS system is not only suited for classic microlensing.   It detects
any object which increases its brightness in a  continuous way. This way
every strange object which changes  its brightness after a long period
of non-variability may be easily noticed.  For example, the possible
double microlens,  OGLE\#7$^4$, which exhibited a very strange light
variations,  would have been detected during the event had the EWS been
implemented in the 1993 season. It should be stressed here that any
object changing its brightness after a long period (years) of remaining
at constant level  is  potentially interesting and worth observing.

Up to now, only bright and obvious objects like novae or  supernovae
were detectable in real time, and still during the event.   With systems
like the EWS it is possible to register any on-going event which is
characterized by a temporary change of its brightness.  As mentioned
above, the EWS can easily be tuned for objects which fade (eg. R~Coronae
Borealis stars).  We also plan to extend the EWS to make it possible to
detect in real time the objects which brighten but which are normally
below the detection limit of the OGLE photometry.

\vskip 3cm

\noindent
Acknowledgements. The OGLE project is supported by the US NFS and Polish
KBN grants.

\vfill
\break

\centerline{R~E~F~E~R~E~N~C~E~S}
\noindent
1. Udalski, A., Szyma\'nski, M., Ka\l u\.zny, J., Kubiak,
M., Krzemi\'nski, W., Mateo, M., Preston, G. W., \& Paczy\'nski, B.
{\it Acta Astronomica}, {\bf 43}, 289--294 (1993).

\noindent
2. Udalski, A., Szyma\'nski, M., Ka\l u\.zny, J., Kubiak, M.,
Mateo, M., \&  Krzemi\'nski, W.  {\it Astrophys J. Letters}, {\bf 426},
L69--L72 (1994).

\noindent
3. Udalski, A., Szyma\'nski, M., Stanek, K. Z., Ka\l u\.zny, J.,
Kubiak, M., Mateo, M., Krzemi{\'n}ski, W., Paczy\'nski, B., and Venkat,
R. {\it Acta Astronomica}, {\bf 44}, 165-189 (1994).

\noindent
4. Udalski, A., Szyma\'nski, M., Mao, S., Di Stefano, R., Ka\l u\.zny,
J.,  Kubiak, M., Mateo, M., and Krzemi\'nski, W.  {\it Astrophys J.
Letters}, (submitted).

\noindent
5. Alcock, C.  {\it et al.}, {\it Nature}, {\bf 365}, 621--623 (1993).

\noindent
6. Alcock, C. {\it et al.}, {\it Astrophys J. Letters}, (submitted).

\noindent
7. Aubourg, E. {\it et al.}, {\it Nature}, {\bf 365}, 623--625 (1993).

\noindent
8. Paczy\'nski, B. {\it Apstrophys. J.}, {\bf 304}, 1--5 (1986).

\noindent
9. Paczy\'nski, B. {\it Apstrophys. J.}, {\bf 371}, L63--L67 (1991).

\noindent
10. Mao, S., \& Paczy\'nski, B. {\it Apstrophys. J. Letters}, {\bf 374},
L37--L40 (1991).

\noindent
11. Paczy\'nski, B. {\it IAU Circular},~\#5997 (1994).

\noindent
12. Schechter, P.L., Mateo, M., and Saha A. {\it Publs astr.\ Soc.\
Pacif.}, {\bf 105}, 1342--1353 (1993).

\noindent
13. Udalski, A., Szyma\'nski, M., Ka\l u\.zny, J., Kubiak, M., \& Mateo,
M. {\it Acta Astronomica}, {\bf 42}, 253--284 (1992).

\vfill
\break
\centerline{Figure Captions}
\noindent
Fig.\ ~1. $66''\times 66''$ fragment of $I$-band CCD frame around
OGLE~\#11. Arrows point to the lensed star.

\noindent
Fig.\ ~2. The light curve of OGLE~\#11 in 1992 -- 1994 observing
seasons. Error bars correspond to estimated 1-$\sigma$ errors. Thick
solid line shows the constant light level from 1992-93 seasons, and
dashed line -- the EWS threshold. Vertical arrow points to the time of
discovery. Thin solid  line shows the best fit of the theoretical
microlensing light curve.

\vfill

\end
\bye